\documentclass{PoS}
\usepackage{epsfig,amsmath,
subfigure,rotating}
\newcommand{\PANDA}{$\overline{\textrm{P}}\textrm{ANDA}$}

\title{Hadrons with $c-s$ content: past, present and future}

\ShortTitle{Hadrons with $c-s$ content}

\author{\speaker{Elisabetta Prencipe} 
        on behalf of the \PANDA~Collaboration\\
        J\"ulich Forschungszentrum\\
        E-mail: \email{e.prencipe@fz-juelich.de}}

\abstract{The \PANDA~detector at FAIR aims to conduct an antiproton-proton experiment with a very high rate capability. It is expected to feature high mass resolution, more than 20 times better than achieved at B-factories. \PANDA~is in a unique position to perform highly resolved mass scan, and to measure the width of very narrow charm and charmonium-like states, whose nature is still unknown, 12 years after their discovery. In this report, we present a method to determine the width of the $D_{s0}^*(2317)^+$. We discuss the future perspectives of \PANDA, based on our present simulations, in relation with the recent measurements performed by LHCb and the performances in this field at the B factories.}

\FullConference{53rd International Winter Meeting on Nuclear Physics,\\
		26-30 January 2015\\
		Bormio, Italy}

\begin{document}
%\begin{linenumbers}

\section{Introduction}
The sector of Charm and Charmonium physics is richer as expected according to the potential model predictions. New resonant states with quite unusual properties have been observed. Prominent examples are the $X(3872)$ and the charged $Z_c^\pm(3900)$ in the Charmonium sector, and the $D_s$ mesons below the $DK$ threshold in the open-charm sector. 

$Strangeness$, in combination with open and hidden Charm, is a topic still to be exploited, in both Charm and Charmonium field. Composite systems of heavy-light quarks have gained the attention of the Charm community since the discovery of the $D_{s0}^*(2317)^+$, that was surprisingly found more than 100 MeV/c$^{\rm 2}$ below the potential model predictions. As a consequence, new theoretical interpretations have been proposed to explain this resonant state, like hadro-charmonia, hybrids, tetraquarks and hadronic molecules. High quality calculations and measurements of the properties of these states are compulsory to decide among the various scenarios, and conclude on their nature. 

We report here on a method to determine the width of the $D_{s0}^*(2317)^+$ with the \PANDA~experiment, proposing a tight mass scan in 100 keV/c$^{\rm 2}$ steps.

\section{Motivation}
Potential models give reliable predictions of $D$ and $D_s$ masses~\cite{godfrey, dipierro}. These particles are bound systems of a heavy quark ($c$), and a light quark ($u$, $d$ or $s$). The $s$-quark represents a ``transition'' case between the sector of light quarks (e.g. $u$ and $d$), and heavy quarks (e.g., $c$, $b$, or $t$). Bound systems composed by a $c$-quark and a $s$-quark, e.g. the so-called $D_s$ mesons, are charged states. On the other hand, $D$ mesons can be neutral and charged.

The $D_{s0}^*(2317)^+$ was observed by BaBar in 2003~\cite{antimo}, at a mass value more than 100 MeV/c$^{\rm 2}$ below the potential model predictions, and it was observed only in the decay to $D_s^+ \pi^0$. To make the picture more complicated, the $D_{s1}(2460)^{+}$ was observed in 2004~\cite{ds2460}. Also in this case, the presence of a spin 1$^+$ state was predicted, but its mass was found 80 MeV below the potential model prediction.  This is not understood, because the $s$-quark could be considered as light up to a certain extent, and the $c$-quark is massive enough, so the perturbative calculations are expected to deliver reliable predictions. Therefore, several alternative theoretical interpretations have been proposed to explain this, and other $D_s^{(*)}$ resonant states. The interpretation of the $D_{s0}^*(2317)^+$ and the  $D_{s1}(2460)^{+}$ as hadro-charmonia~\cite{hadroc}, hybrids~\cite{tetra}, or pure $c \bar s$ states~\cite{purecs} have been suggested.

The $D_{s0}^*(2317)^+$ width is nowadays unknown: an upper limit of 3.8 MeV has been set~\cite{pdg} to the width measurement of this very narrow state by experiments at B factories. The same stands for the  $D_{s1}(2460)^{+}$. In the recent theoretical work of Ref.~\cite{hanhart}, the $D_{s0}^*(2317)^+$ width is evaluated under the hypothesis that it is a pure $c \bar s$ state, and under the hypothesis of being a molecular state. 
The former case leads to a width of $\sim$30 keV, whereas in the latter case the width is predicted to be $\sim$133 keV. In this theoretical paper, there are arguments in favor of the molecular nature of the $D_{s0}^*(2317)^+$. The two hypotheses differ basically because of the contribution of the $D^+K^0$ and $D^0K^+$ loops to the rate of the process, which gives rise to an additional contribution of the hadronic width of the $D_{s0}^*(2317)^+$ of about 100 keV. To unambiguously conclude on the preferred molecular description of the state, a new generation of experiments able to scan the mass of the resonant state in $\sim$100~keV/c$^{\rm 2}$ steps is needed. 

\begin{figure}[here] 
\begin{center}
\mbox{
\subfigure{\scalebox{0.4}{\includegraphics{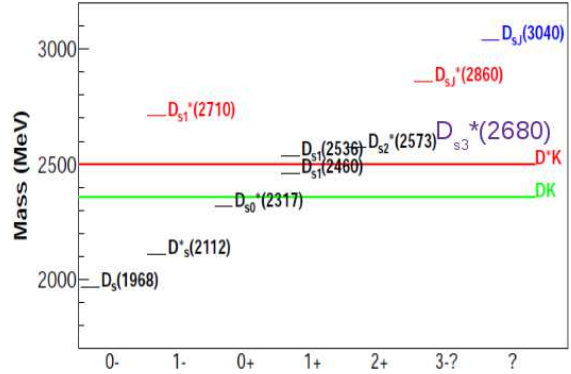}}}}
\caption{\label{fig1} $c \bar s$ mass spectrum as function of the spin J$^P$, today. The $D_s(2860)$ has been recently interpreted as an admixture of spin 1 and spin 3 states by LHCb~\cite{lhcb2860}.} 
\end{center}
\end{figure}

In a $\bar p p$ process, it is possible to reach such a high mass resolution only with a very high  beam momentum resolution. In this situation, the $D_s$ mesons can be produced as charged pairs. So, the process to be investigated could be  $\bar p p \rightarrow D_s^-  D_{s0}^*(2317)^+$, where $ D_s^-$ is the ground state of the $c \bar s$ spectrum. In Fig.~\ref{fig1}, the $c \bar s$ spectrum is reported, populated by many states, mainly discovered in the past 12 years. The minimum antiproton beam momentum that one needs to run, in order to produce the $D_s^-  D_{s0}^*(2317)^+$ pair via $\bar p p $ annihilation, is $p_{beam}$ = 8.80235 GeV/c. Here, a mass resolution of 100 keV/c$^{\rm 2}$ can be reached only if the beam momentum resolution is $\sim$10$^{-5}$. An experiment with these features does not exist, presently. However, the future \PANDA~experiment~\cite{PB} will reach a beam momentum resolution $\Delta p/ p$ of the order of 10$^{-5}$ (in the high resolution mode), which is ideally what is needed to perform  the challenging measurement of the width of the  $D_{s0}^*(2317)^+$  and  the $D_{s1}(2460)^+$. 
A crucial point is of course the amount of luminosity that is needed for such a measurement. This number depends on the cross section of the  $\bar p p \rightarrow D_s^-  D_{sJ}^{*+}$ process, which is still unknown: $\cal L \cdot \sigma \cdot \epsilon$ = $N$, where $\cal L$ is the integrated luminosity, $\sigma$ is the cross section of the process, $\epsilon$ is the reconstruction efficiency for the resonant state, and $N$ is the number of fitted $D_s^-  D_{sJ}^{*+}$ pairs. 

Several theoretical studies have been performed in the past to estimate the cross section of $\bar p p$ to open-charm. They have exploited different theoretical approaches (perturbative QCD~\cite{braaten, mannel}, baryon exchange~\cite{krein} or quark model~\cite{faustov}, (un)quenched lattice QCD  calculations), leading to different conclusions. 

In Fig.~\ref{fig7}, the resultant cross section is estimated to be in the range [1;10] nb (quark model), and [20;30] nb (baryon exchange model), as shown, respectively.  The results of these models are based on SU(4) symmetry, which are expected to be only valid for $D$($D_s$) ground states~\cite{krein}. For excited $D$ or $D_s$ states, the situation is more complicated. One cannot rely on predictions on the coupling constant evolution depending on the new energy scale. Perturbative calculations do not work in this case. One would need to look into a (semi)inclusive process, in which the charm quark has to hadronize into $D_{s0}^*(2317)^+$. Not much is known so far about the fragmentation functions for this state. 

In the most conservative quark model approach, the $D_{s0}^*(2317)^+$, the   $D_{s1}(2460)^+$ and the $D_s(2860)^+$ are the only anomalies, in the $c \bar s$ spectrum. It is interesting to note that the $D_s(2860)^+$ is observed above the $D^*K$ threshold, with a larger width compared to that of  the very narrow $D_{s0}^*(2317)^+$ and $D_{s1}(2460)^+$, observed below the $D^*K$ threshold.  Lattice QCD calculations (for example, as reported in Refs.~\cite{bali, another}) suggest that $DK$ scattering amplitudes are required to obtain the large mass shifts of 180 MeV/c$^{\rm 2}$ for the $D_{s0}^*(2317)^+$. A study of the $B_s$ counterpart of these $D_s$ states would help to understand their nature. However, $B_s$ decays will be not part of the \PANDA~physics program, because  the energy in the center of mass system that \PANDA~can reach is limited to 5.5 GeV/c$^{\rm 2}$.

Regge trajectories are also used in the cross section calculations, for example in the process of $\bar p p$ to baryons. However, when developing calculations up to higher orders, divergences occur that are difficult to cure. Therefore, it is not easy to perform theoretical calculations in  a rigorous way for the process  $\bar p p \rightarrow$ open-charm, especially when excited states are involved. Data are needed, and an experiment providing a high-momentum resolution antiproton beam is crucial to perform these studies.

\section{The \PANDA~approach}

\begin{figure}[htb] 
\begin{center}
\mbox{
\subfigure{\scalebox{0.4}{\includegraphics{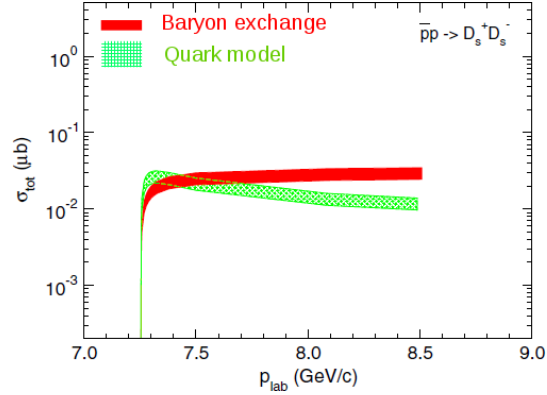}}}}
\caption{\label{fig7} Cross section calculation of the $\bar p p \rightarrow D_s^+ D_s^-$ reaction within the SU(4) symmetry. Results are taken from Ref.~\cite{krein}.} 
\end{center}
\end{figure}

A (semi)inclusive analysis for the channel $\bar p p \rightarrow D_s^- D_{s0}^*(2317)^+$ is proposed. It gives a larger cross section compared to the exclusive analysis. In the absence of a rigorous prediction for the cross section of $\bar p p \rightarrow D_s^- D_{s0}^*(2317)^+$, we may consider  valid for the $\bar p p \rightarrow D_s^-  D_{s0}^*(2317)^+$ process, the same cross section prediction as for $\bar p p \rightarrow D_s^-  D_s^+$ (see Fig.~\ref{fig7}). The calculation reported in Ref.~\cite{krein} extends up to $p$ = 8.5 GeV/c, and the antiproton beam momentum, which is needed for this analysis, is $p_{beam}~ \geq$8.8 GeV/c.
We thus extrapolate our results under the assumption that the cross section of the $\bar p p \rightarrow D_s^-  D_{s0}^*(2317)^+$ process is in the range [20;30] nb. 
\subsection{Experimental overview}
\begin{figure}[ht] 
\begin{center}
\mbox{
\subfigure[]{\scalebox{0.48}{\includegraphics{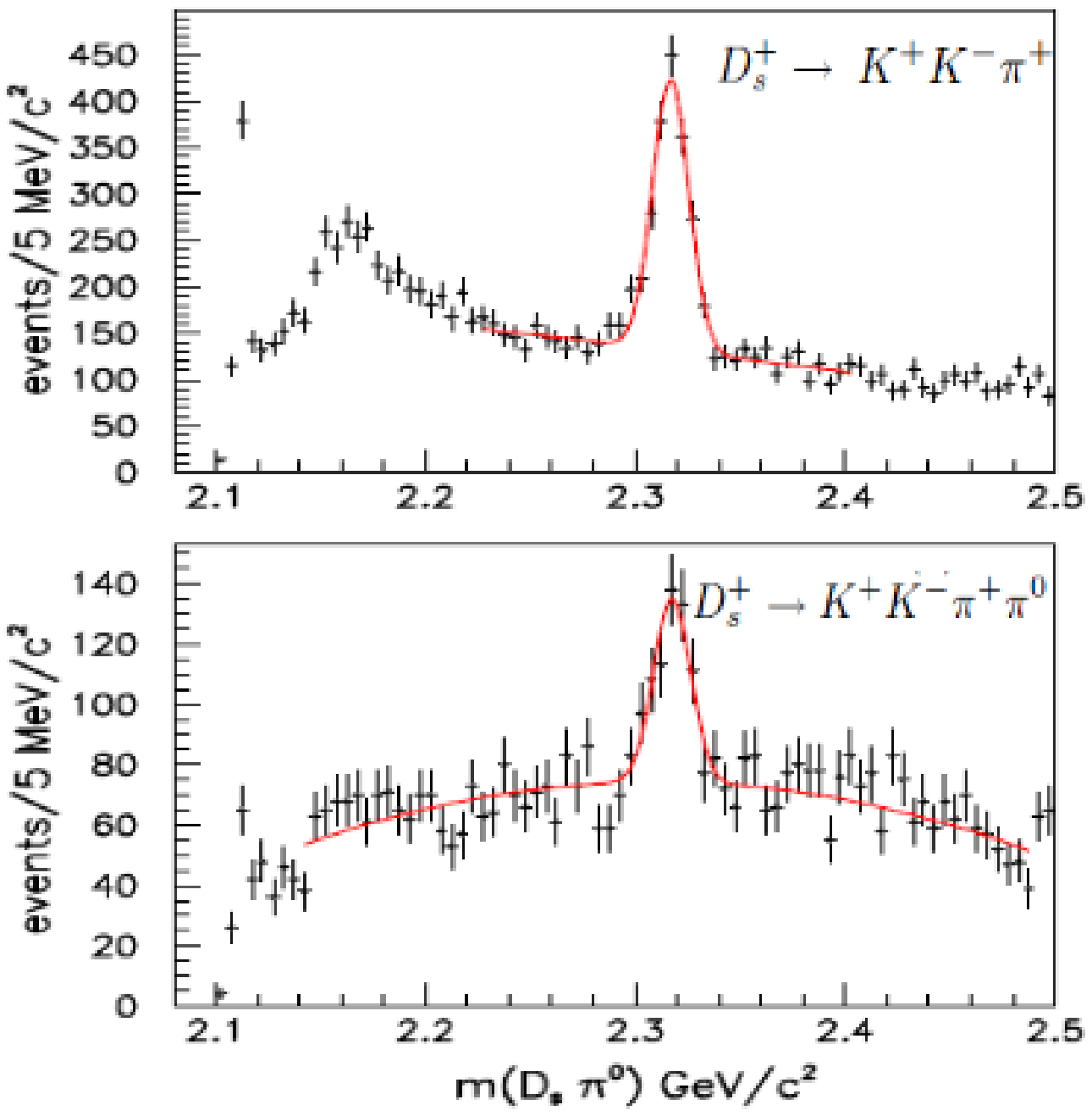}}}\quad
\subfigure[]{\scalebox{0.49}{\includegraphics{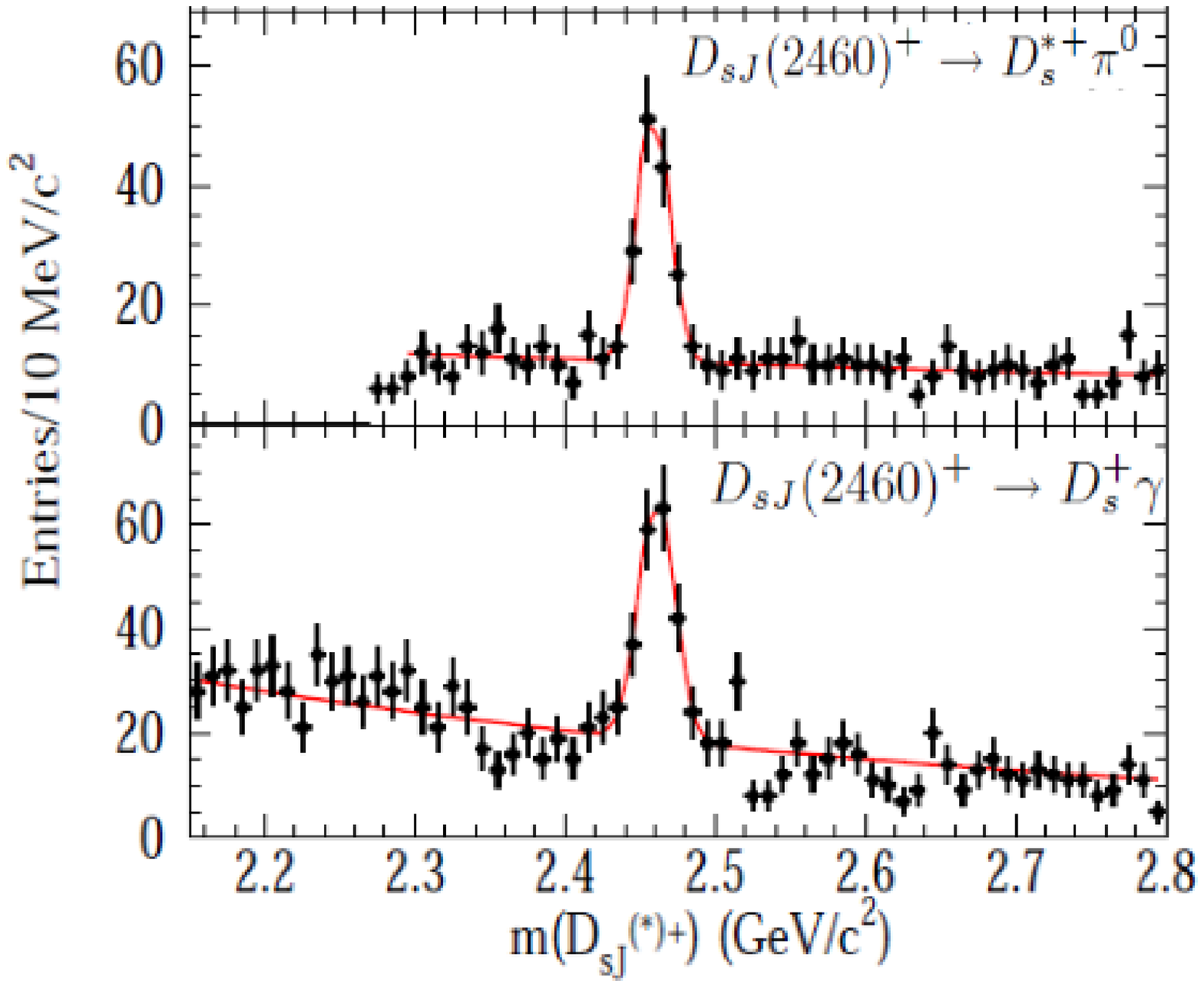}}}
}
\caption{\label{fig8} (a) $D_S^+ \pi^0$ invariant mass system, in the inclusive analysis of $e^+ e^- \rightarrow \bar c c$ at the energy near 10.6 GeV (BaBar data)~\cite{antimo}: a clear peak at the mass M = 2.371 GeV/c$^{\rm 2}$ is observed. (b) Observation of $D_{s1}(2460)^+$  at BaBar in two different decay modes~\cite{ds2460}.} 
\end{center}
\end{figure}

 Figure~\ref{fig8} shows the published results from the BaBar experiment~\cite{antimo, ds2460}. Clear peaks have been observed, corresponding to the $D_{s0}^*(2317)^+$ and  the $D_{s1}(2460)^+$, respectively. A structure in the $D_s^+ \pi^0$ invariant mass, corresponding to the  $D_{s0}^*(2317)^+$, is supposed to have $J^P$ = 0$^+$.  The search for other decays modes led to the discovery of an additional state, the  $D_{s1}(2460)^+$. This new state was seen by analyzing the invariant mass of $D_s^+ \gamma$ and $D_s^{*+} \pi^0$. Since the  $D_{s1}(2460)^+$ was not observed decaying to  $D_s^+ \pi^0$, we can conclude that it is not a spin-0 resonance, but it has spin 1. The angular analysis performed by BaBar favors this spin assignment.

If  $D_{s0}^*(2317)^+$ and $D_{s1}(2460)^+$ are the  $J^P$ = 0$^+$  and $J^P$ = 1$^+$ states, respectively, belonging to the same family in the $c \bar s$ spectrum, they are expected to decay to the channels summarized in Table~\ref{tab1}. However, some of the decays are not observed. This disfavors the interpretation of the  $D_{s0}^*(2317)^+$ and the $D_{s1}(2460)^+$ as the missing $J^P$ = 0$^+$  and $J^P$ = 1$^+$ states of the $c \bar s$ spectrum. In order to understand what these states are, more searches were performed.  Their present status is summarized in Fig.~\ref{fig1}, where many excited $D_s$ states are shown, even above the $DK$ threshold.

\begin{table}[htb]
\caption{ \label{tab1} Summary of decay channels for the $D_{s0}^*(2317)^+$ and the $D_{s1}(2460)^+$. }
\begin{center}
\begin{tabular}{lrcccl}
 \hline \hline
Decay channel & $D_{s0}^*(2317)^+$ & $D_{s1}(2460)^+$\\ \hline
 $D_s^+ \pi^0$  & Seen& Forbidden\\ 
 $D_s^+ \gamma$ & Forbidden& Seen \\ 
 $D_s^+ \pi^0 \gamma$ non resonant &Allowed & Allowed\\
$D_s^* (2112)^+  \pi^0$ & Forbidden & Seen \\
$D_{s0}^* (2317)^+  \gamma$ & $-$ & Seen \\
$D_s^+ \pi^0 \pi^0$ & Forbidden & Allowed \\ 
$D_s^+ \gamma \gamma$ (non resonant)& Allowed & Allowed\\ 
$D_s^* (2112)^+  \gamma$ & Allowed & Allowed \\
$D_s^+ \pi^+ \pi^-$  & Forbidden& Seen\\ \hline \hline
\end{tabular}
\end{center}
\end{table}

 From past experiments, we know that a good technique for measuring the width of a narrow state is to scan its mass. With this technique,  and with a beam momentum resolution of $\sim$500~keV/c, the experiment E760, for instance, measured the width of the $J/\psi$: $\Gamma$ = (99 $\pm$ 12 $\pm$ 6)~keV~\cite{e760}. For comparison, \PANDA~ is designed to have a beam momentum resolution $\leq$100~keV/c.

\PANDA~ has multiple interests in analyzing  the $\bar p p \rightarrow D_s^-  D_{s0}^*(2317)^+$ channel, namely:
\begin{itemize}
\item a fit to the excitation function of the cross section, to extract the width ($\Gamma$) of the $D_{s0}^*(2317)^+$ and $D_{s1}(2460)^+$;\\
\item  the production cross section determination;\\
\item  a chiral symmetry breaking test;\\
\item  the study of mixing of $D_{sJ}$ states with same spin and parity;\\
\item  study of the invariant mass system of $D_s D_{sJ}^*$, and search for 4-quark states with $strange$ content in the Charmonium field. 
\end{itemize}
In this report we only discuss the first two items. The next section reports the status of the \PANDA~simulations.

\subsection{Strategy}
\label{strate}
\begin{figure}[ht] 
\begin{center}
\mbox{
\subfigure{\scalebox{0.35}{\includegraphics{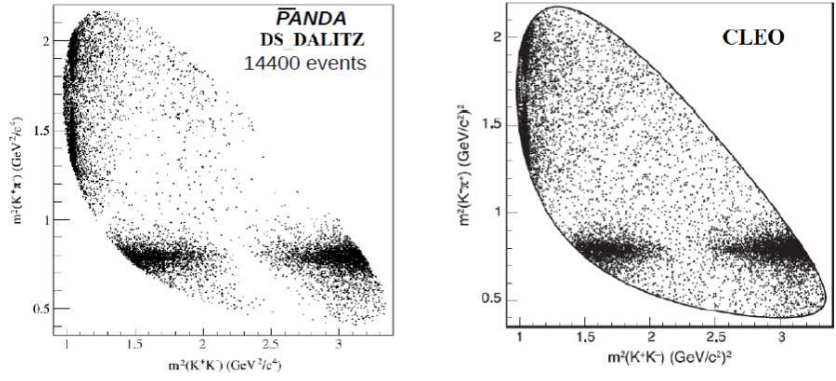}}}
}
\caption{\label{fig9} Simulation in Pandaroot for the reconstruction of $D_s^+ \rightarrow K^+ K^- \pi^+$ in the process $\bar p p \rightarrow D_s^+ D_s^-$: 14400 events have been generated using the DS-DALITZ model in EvtGen (left); Dalitz plot based on real data from the CLEO experiment for $D_s^+ \rightarrow K^+ K^- \pi^+$ (right)~\cite{cleo}.} 
\end{center}
\end{figure}

In \PANDA, the framework used to simulate $\bar p p \rightarrow D_s^-  D_{s0}^*(2317)^+$ events is PandaRoot~\cite{pandaroot}. It is based on Virtual Monte Carlo (VMC), that makes use of  $Geant3$~\cite{g3} for this specific analysis. A full simulation is performed, with a detailed magnetic B field map (constant B = 2 T in the central part; B = 2 T$\cdot$m in the dipole area). 
Signal events are simulated using the Monte Carlo (MC) generator EvtGen~\cite{evtgen}; background events are simulated using the Dual Parton Model (DPM)~\cite{dpm}. The reconstruction chain under study is: $\bar p p \rightarrow D_s^-  D_{s0}^*(2317)^+$, $D_s^- \rightarrow K^+ K^- \pi^-$. In an exclusive reconstruction process, we would have: $ D_{s0}^*(2317)^+ \rightarrow D_s^+ \pi^0$, $\pi^0 \rightarrow \gamma \gamma$. The model used to simulate $D_s$ events is the so called DS-DALITZ, which reproduces all internal structures of the $K \pi$ and $KK$ systems in the $D_s^+ \rightarrow K^+ K^- \pi^+$ decay. The DS-DALITZ model is part of EvtGen, and it is based on the BaBar/CLEO data (see Fig.~\ref{fig9}).

At the production threshold of the $D_s^-  D_{s0}^*(2317)^+$ system in $\bar p p$ annihilation, the antiproton beam is set to $p_{beam}$ = 8.80235 GeV/c, corresponding to an invariant mass of the system M$_{tot}$ = 4.286 GeV/c$^{\rm 2}$. Particle identification is used to separate kaons from pions. A cut on the photon momentum ($p_\gamma$ >50 MeV/c) and track momentum ($p_{track}$ > 100 MeV/c) is applied at the level of an event pre-selection. This reduces the background due to very low momentum tracks. To improve the mass resolution and efficiency, an inclusive study in a single-tag mode is performed. This implies that we fully reconstruct the $D_s$ meson via the detected decay particles, and optimize a dedicated selection. We thus  obtain the $ D_{s0}^*(2317)^+$ as missing mass of the event  $\bar p p \rightarrow D_s^-  D_{s0}^*(2317)^+$.

\begin{figure}[ht] 
\begin{center}
\mbox{
\subfigure[]{\scalebox{0.40}{\includegraphics{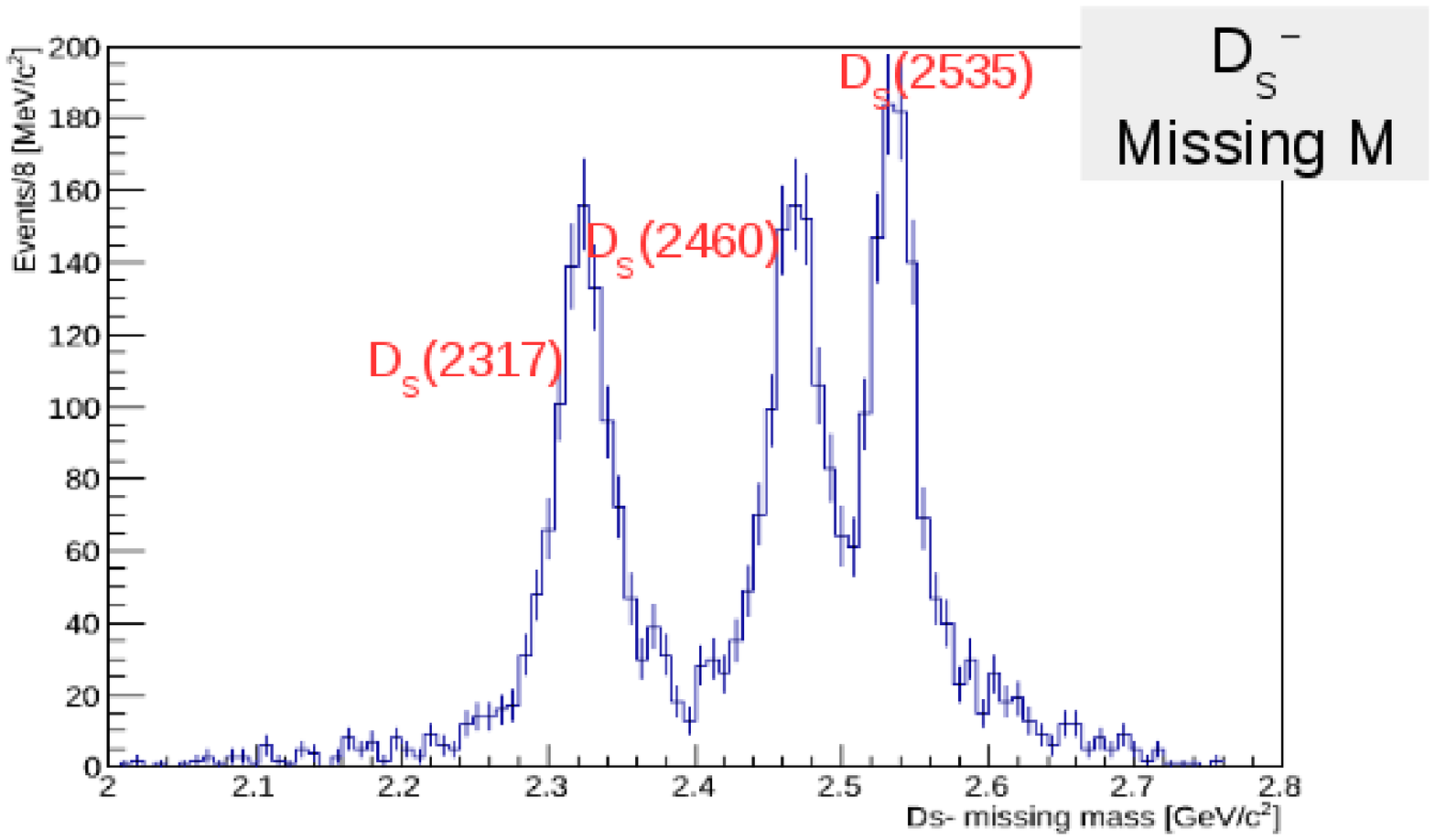}}}\quad
\subfigure[]{\scalebox{0.41}{\includegraphics{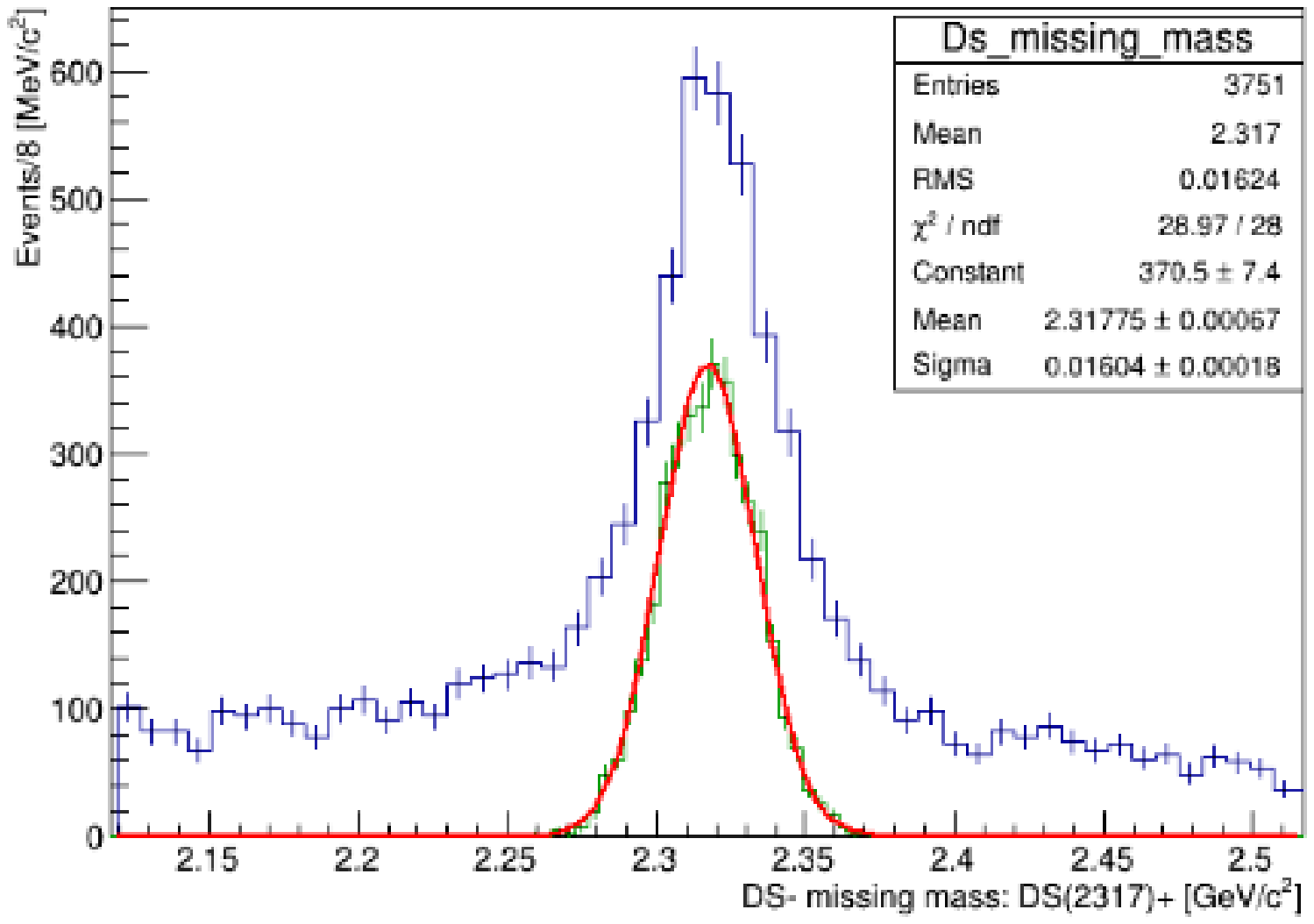}}}
}
\caption{\label{fig10} (a) Mass values of several $D_{sJ}$, in an inclusive analysis of the channel $\bar p p \rightarrow D_s^- D_{sJ}^{(*)+}$, as indicated. (b) Inclusive reconstruction of $ D_{s0}^*(2317)^+$ in the process $\bar p p \rightarrow D_s^- D_{s0}^*(2317)^+$. The blue curve represents signal + combinatorial background; the red curve fits the signal shape (green distribution). For this analysis, 10 000 events have been simulated.} 
\end{center}
\end{figure}

We perform two studies in parallel: an inclusive study including 3 different $D_s^{(*)}$ states, as reported in Fig.~\ref{fig10}(a), and a specific simulation where only the excited $D_s$ state is the $ D_{s0}^*(2317)^+$ (see Fig.~\ref{fig10}(b)). In both figures, only the pre-selection has been applied, and we select $\sim$30$\%$ of $D_s$ on a sample of 10 000 generated events. A mass resolution of $\sim$16 MeV/c$^{\rm 2}$ is obtained by a Gaussian fit. The background cross section is evaluated to be in the order of $\sim$2.2 mb, which has to be compared with a signal cross section of the order of  a few nb. The rejection of the high combinatorial background is one of the main challenges of this analysis.

\subsection{A dedicated selection for $\bar p p \rightarrow D_s^-  D_{s0}^*(2317)^+$ reaction}

Once the pre-selection is fixed, a dedicated selection is studied. It involves kinematic variables, such as the cosine of the  angles among the $D_s$ daughters, the $p_{t,z}$ momentum resolution, the $D_s$ decay vertex position, and  $\Delta E$. The variable $\Delta E$ is defined as the difference between the measured energy in the center of mass system, and its expected value. It can be parameterized by a double Gaussian function for signal events, centred around zero, and a polynomial function for background events. 

\begin{figure}[ht] 
\begin{center}
\mbox{
\subfigure[]{\scalebox{0.40}{\includegraphics{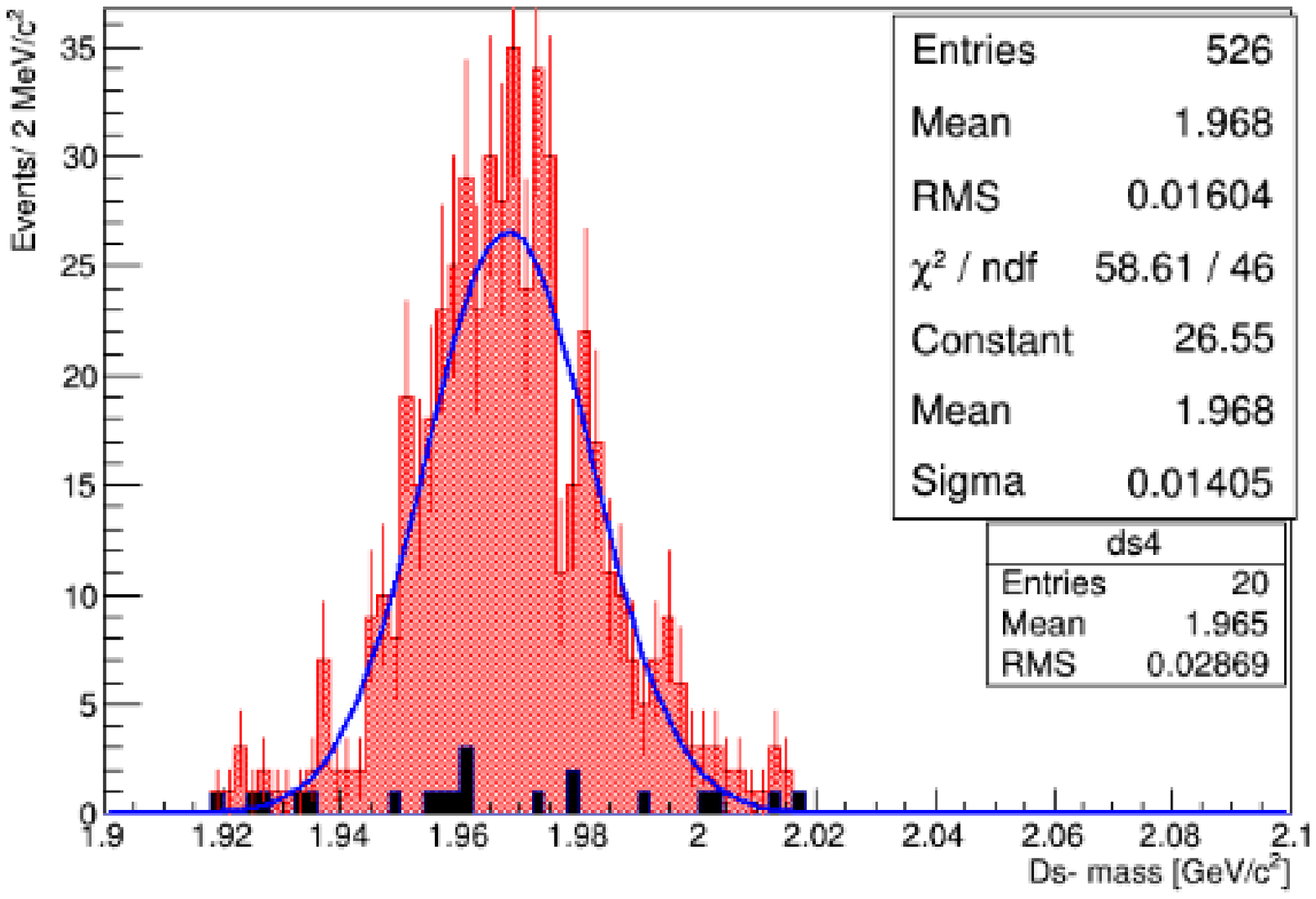}}}\quad
\subfigure[]{\scalebox{0.45}{\includegraphics{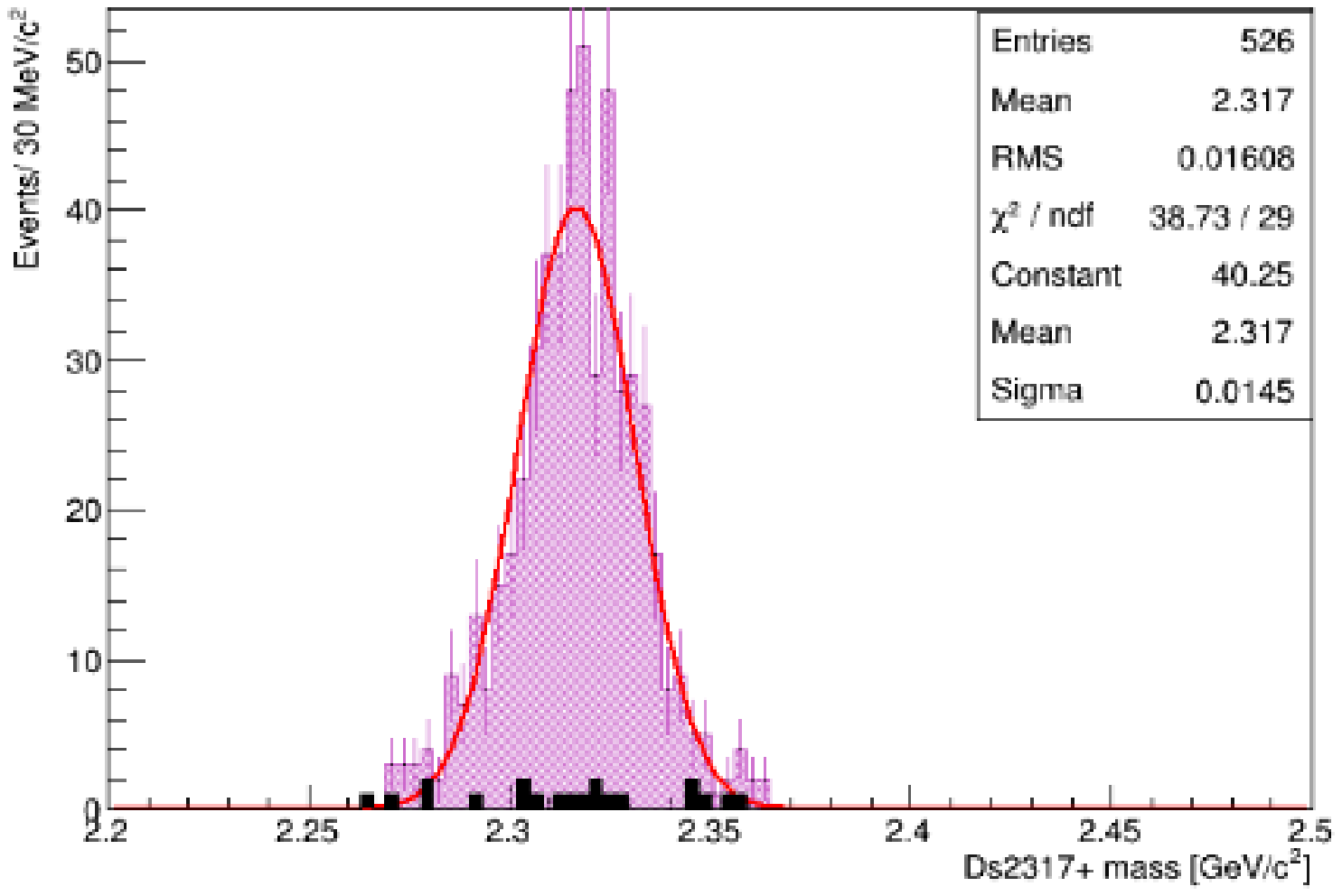}}}
}
\caption{\label{fig13} (a) The reconstructed $D_s$ mass refers to the $K^+K^-\pi^-$ system. Comparison between signal (red) and  background (black), after a dedicated selection. (b) The $D_{s0}^*(2317)^+$ mass refers to the missing mass of the event in the $\bar p p \rightarrow D_s^- D_{s0}^*(2317)^+$ process. Comparison between signal (violet) and background (black), after a dedicated selection. Under the hypothesis of a signal cross section equal to 20 nb, the background sample has to be scaled by a factor 60.} 
\end{center}
\end{figure}

Figure~\ref{fig13} shows the results of this preliminary selection. The  $D_{s0}^*(2317)^+$ is evaluated as missing mass of the event. Moreover, simulation studies show that the mass resolution of the $D_s^-  D_{s0}^*(2317)^+$ system is expected to be significantly smaller than that of the reconstructed $D_s^-$ and the $D_{s0}^*(2317)^+$, shown in Fig.~\ref{fig13}(a) and Fig.~\ref{fig13}(b), respectively.

\begin{figure}[ht] 
\begin{center}
\mbox{
\subfigure{\scalebox{0.49}{\includegraphics{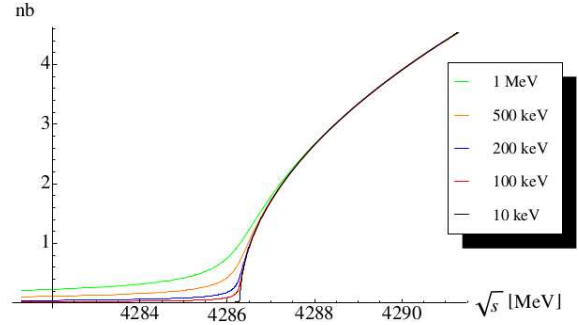}}}}
\caption{\label{fig15} Excitation function of the cross section of the process $\bar p p \rightarrow D_s^- D_{s0}^*(2317)^+$, as function of the energy in the center of mass system of $D_s^- D_{s0}^*(2317)^+$. The plot is taken from Ref.~\cite{marius}.}
\end{center}
\end{figure}

We plan to scan the invariant mass system of $D_s^- D_{s0}^*(2317)^+$ in 100 keV/c$^{\rm 2}$ steps, at 15 energy-scan points. We need to collect many points as the  $ D_{s0}^*(2317)^+$ line shape is not known at all. We will extract the excitation function of the cross section as function of the energy; at the production theshold of the $D_s^- D_{s0}^*(2317)^+$ pair, the excitation function of the cross section depends on the mass and width of the resonant state. In Fig.~\ref{fig15} this curve is shown, for different input values of the width~\cite{marius}.

With the preliminary selection described in this report, we obtain a reconstruction efficiency of $\sim$17.5$\%$. Assuming that \PANDA~will run in high resolution mode (e.g. $\cal L$ = 0.86 pb$^{-1}$/day), for which we tag the $D_s^- \rightarrow K^+K^- \pi^-$ (BR ($D_s^- \rightarrow K^+K^- \pi^-) \sim$6$\%$), in the hypothesis to run 3000 events per scan point, we will need from 7 up to 11 days of data taking per scan point, on the signal cross section assumed to be in the range of [20;30] nb, as supported by Ref.~\cite{krein}. In this situation, the ratio of signal (S) over background (B) events is S/B = 1/28. Before appling this preliminary selection, S/B = 1/10$^{\rm 6}$.

For comparison, in the first 3 years of data taking, the BaBar experiment recorded 1267 event yield. Therefore, we conclude that with \PANDA~the reconstruction efficiency of the  $D_{s0}^*(2317)^+$ is significantly higher than at the B factories. The performance of the proposed inclusive analysis, based on PandaRoot MC simulations, looks therefore promising.   
\section{Summary}
Charm physics is a field of high interest, in which several questions are still unsolved. Among these, the measurement of the width of the narrow $D_s$ states below the DK threshold is important to fully understand the $c \bar s$ spectrum. \PANDA~offers a unique opportunity to perform this measurement, and the results of simulations with PandaRoot are very promising. This measurement cannot be performed  in the early stage of \PANDA~data taking. The cross section of the process $\bar p p \rightarrow D_s^- D_{s0}^*(2317)^+$ could be estimated to be in the range [20;30] nb (as explained in Ref.~\cite{krein}), thus in the worse scenario, corresponding to a cross section of 20 nb, we would need of about 165 days to perform the full scan of the  $D_s^- D_{s0}^*(2317)^+$ system.  The results that we aim for, will reach  a level of precision never achieved before. Therefore, we consider this analysis a highlight of the \PANDA~physics program. The optimization of the selection, and the numbers quoted in this report, are work in progress. We plan to complete the first full simulation campaign for the analysis of $\bar p p \rightarrow D_s^- D_{s0}^*(2317)^+$ by the summer 2015.
%\end{linenumbers}

\end{document}